\documentclass[twocolumn]{article}
\pdfoutput=1
\usepackage[utf8]{inputenc}
\usepackage{placeins}
\usepackage{graphicx}
\usepackage{nopageno}
\usepackage{url}
\usepackage[breaklinks]{hyperref}

\hypersetup{
    colorlinks=true,
    linkcolor=blue,
    filecolor=magenta,      
    urlcolor=blue,
    citecolor=red,
    breaklinks=true,            
}
\usepackage{breakurl}

\providecommand{\keywords}[1]
{
  \small	
  \textbf{\textit{Keywords---}} #1
}

\title{COVID-19: A Survey On Public Medical Imaging Data Resources}
\author{
    Roman Kalkreuth\\
    Department of Compuer Science\\
    TU Dortmund University\\
    roman.kalkreuth@tu-dortmund.de\\
    \url{http://www.cs.tu-dortmund.de}
  \and
    Paul Kaufmann\\
    Chair for Computational Intelligence\\
    Johannes Gutenberg University Mainz\\
    paul.kaufmann@uni-mainz.de\\
    \url{https://ci.bwl.uni-mainz.de/}
}
\date{\today}

\date{Last Update: \today. Version 2.0}

\begin{document}

\maketitle

\begin{abstract}
This regularly updated survey provides an overview of public resources that offer medical images and metadata of COVID-19 cases. The purpose of this survey is to simplify the access to open COVID-19 image data resources for all scientists currently working on the coronavirus crisis. 
\end{abstract}

\vspace{4mm}

\vspace{-2mm}

\section{Introduction}
Recent study results have show that medical imaging can be used for the diagnosis of COVID-19~\cite{li2020coronavirus}. Moreover, two recent studies indicate that COVID-19 is possibly better diagnosed with radiological imaging when compared to the PCR-Test~\cite{ai2020correlation,fang2020sensitivity}. To classify computer tomography (CT) images of COVID-19 cases automatically, a Convolutional Neural Network has been used in~\cite{xu2020deep}. The  study aimed to answer the questions, if artificial intelligence technology can be used to early screen COVID-19 patients from their CT images and what is the corresponding diagnostic accuracy. The study also focused on the development of an early screening model to distinguish COVID-19 pneumonia from Influenza-A viral pneumonia and healthy cases. The experiments showed that the overall accuracy was 86.7\% from the perspective of CT cases as a whole. 

Recently, a call has been launched for scientists working in the field of artificial intelligence to focus their research and knowledge on the current coronavirus crisis~\cite{ai_call}. At the moment, there is a lack of sufficiently large and publicly available records of image data. Data is, however, essential for machine learning techniques. With the intention to simplify the access to currently publicly available image data and metadata we started this survey. 

\vspace{-2mm}

\section{Motivation}
\vspace{-1mm}
At present, medical capacities are severely limited in many countries. Also, the flow of information and exchange between physicians and scientists is limited due to the lockdowns in various countries. More massive data sets with image data of COVID-19 cases are, therefore, either not available or not publicly accessible. However, since these data sets are urgently needed to support the diagnosis and treatment of COVID-19 with machine learning or artificial intelligence technologies, we would like to contribute with this initiative to the full exploration and further development of the currently available potential of different technologies.

\vspace{-2mm}

\section{About the Initiative}
\vspace{-1mm}
Since the authors are currently working on the detection of COVID-19 with machine learning themselves, this survey will be updated regularly. We would be very grateful if members of the community could share sources with us, which are not yet listed here. Please report all errors to our email addresses. We will fix them as soon as possible.  With this initiative, we also would like to encourage all scientists to make their data sets publicly available to the community, provided that data protection allows it. This initiative can also be supported by making it better known with a citation.

\newpage

\FloatBarrier

\begin{table*}
\begin{center}
\rotatebox{90}{
\scalebox{0.8}{
\begin{tabular}{| c | c | c | c | c | c | c | c |}
\hline
\textbf{Resource/Dataset} & \multicolumn{6}{ c |}{\textbf{Types of Data}} & \textbf{Organization/Platform} \\ 
\cline{2-7}

& \textbf{CT} & \textbf{X-Ray} & \textbf{MRT} & \textbf{Ultrasound}  & \textbf{Metadata} & \textbf{Case review} & \\
 &  &  &  & & & & \\ \hline
 
\href{https://github.com/ieee8023/covid-chestxray-dataset}{COVID-19 Image Data Collection}~\cite{cohen2020covid,cohen_github_covid19} & Y  & Y  & N & N & Y & N &\href{https://reproducibilityinstitute.org/w/}{Institute for Reproducible Research} \\ \hline

\href{https://pages.semanticscholar.org/coronavirus-research}{COVID-19 Open Research Dataset (CORD-19)}~\cite{cord19} & Y  & Y  & U  & U & Y & Y &\href{https://allenai.org/}{Allen Institute for AI and partners} \\  \hline

\href{https://github.com/lindawangg/COVID-Net/}{COVID-Net Open Source Initiative}~\cite{covid-net} & N  & Y  & N & N & N & N & \href{https://uwaterloo.ca/vision-image-processing-lab/}{Vision and Image Processing Lab}\\ \hline

\href{https://github.com/jannisborn/covid19_pocus_ultrasound}{COVID-19 Pocus Ultrasound Dataset}~\cite{born2020pocovidnet,born2020pocovidnet_dataset} & N & N & N & Y & Y & N & \href{https://bsse.ethz.ch/}{Dep. for Biosystems Science and Engineering}\\ \hline

\href{https://www.sirm.org/category/senza-categoria/covid-19/}{SIRM COVID-19 Database}~\cite{sirm} & Y & Y  & U & U & Y & Y & \href{https://www.sirm.org/en/}{Italian Society of Radiology}\\ \hline

\href{https://bsticovid19.cimar.co.uk/}{COVID-19 BSTI Imaging Database}~\cite{bsti}&  Y   & Y & N & U &  Y & Y & \href{https://www.bir.org.uk/}{British Institute of Radiology} \\ \hline

\href{https://www.kaggle.com/tawsifurrahman/covid19-radiography-database}{COVID-19 Chest X-Ray  Database}~\cite{kaggle-covid19-1} & N &  Y & N & N & Y & N & \href{https://www.kaggle.com/}{Kaggle} \\ \hline

\href{https://www.kaggle.com/bachrr/covid-chest-xray
}{COVID-19 Chest X-Ray}~\cite{kaggle-covid19-2} & N &  Y & N & N & Y & N & \href{https://www.kaggle.com/}{Kaggle} \\ \hline

\href{http://www.medicalsegmentation.com/covid19/}{COVID-19 
CT Segmentation Dataset}~\cite{covid19_ct_seg} & Y & N & N & N & N & N & \href{http://medicalsegmentation.com/}{medicalsegmentation.com} \\ \hline

\href{https://github.com/UCSD-AI4H/COVID-CT}{COVID-CT}\cite{covid_ct,zhao2020COVID-CT-Dataset} & Y & N & N & N & Y & N & \href{https://ucsd.edu/}{University of California San Diego}\\ \hline

\href{https://data.mendeley.com/datasets/2fxz4px6d8/4}{Augmented COVID-19 X-ray Images Dataset} \cite{augmented_c19_xray} & N & Y & N & N & N & N & \href{https://hijjawi.yu.edu.jo/index.php/depts/biomedical-engineering}{Dep. of Biomedical Systems \& Informatics Engineering}\\ \hline

\href{https://www.coronacases.org}{coronacases.org} \cite{coronacases.org} & Y & N & N & U & Y & Y&  \href{https://raioss.com/en_US/}{Radiology Artificial Intelligence} \\ \hline

\href{https://www.eurorad.org/}{eurorad.org} \cite{eurorad.org} & Y & Y & U & U & Y & Y & \href{https://www.myesr.org/}{European Society of Radiology} \\ \hline

\href{https://www.radiopaedia.org/}{radiopaedia.org}~ \cite{radiopaedia.org} & Y & Y & U & U & Y & Y & \href{https://radiopaedia.org/}{Radiopaedia}\\ \hline

\end{tabular}}
}
\end{center}
\caption{List of COVID-19 Medical Imaging Resources and Datasets}
\label{covid19_list}
\end{table*}
\FloatBarrier

\section{List of Resources}

Table~\ref{covid19_list} shows a list of resources and datasets which provide publicly accessible medical imaging data of COVID-19 cases. The types of data are classified by CT, X-Ray, magnetic resonance tomography (MRT), metadata of the corresponding patient/case, and case review. The respective classes are annotated with \textbf{Y} for \textit{yes}, \textbf{N} for \textit{no} and \textbf{U} for \textit{unknown}.

\section{Note of the Authors}
We would like to point out not to claim any form of diagnostic performance of models until clinical studies have been performed. 

\bibliographystyle{acm}
\bibliography{references}
\end{document}